# Mechanism for linear preferential attachment in growing networks

Xinping Xu[♀], Feng Liu , Lianshou Liu

Institute of Particle Physics, Huazhong Normal University, Wuhan 430079, China

**Abstract**

The network properties of a graph ensemble subject to the constraints imposed by the expected degree sequence are studied. It is found that the linear preferential attachment is a fundamental rule, as it keeps the maximal entropy in sparse growing networks. This provides theoretical evidence in support of the linear preferential attachment widely exists in real networks and adopted as a crucial assumption in growing network models. Besides, in the sparse limit, we develop a method to calculate the degree correlation and clustering coefficient in our ensemble model, which is suitable for all kinds of sparse networks including the BA model, proposed by Barab´asi and Albert.

[♀] Corresponding author. E-mail address: xuxp@iopp.ccnu.edu.cn

# 1. Introduction

For last few years, interest and works have been concentrated on the small-world phenomenon and scale free behavior observed in real networks [1－5]. The WS model [6] and BA model [7] provide excellent explanations for these features, respectively. Especially, the BA model embodies the evolutional and dynamical process and generates a very rich phenomenology that captures many of the complex features emerging in the analysis of real networks. Much attention has been focused on proposing models to reproduce the observed scale-free distribution basing on the variation of BA model [8－10].

In BA model [7], two ingredients leading to the scale-free distribution are growth and linear preferential attachment. As a matter of fact, the preferential attachment, which indicates that the likelihood of receiving new edges increases with the node's degree, is a central ingredient of all models to generate scale-free networks. Some authors measured the preferential attachment effect in real networks. *e.g.* Jeong, Ne´da, Baraba´si et al[11], have measured the preferential attachment effect in four real networks － the citation network of articles, the Internet, the co-authorship network of researchers, and the actor collaboration network. It is found that the first two networks have linear preferential attachment, and the latter two networks have sublinear preferential attachment. It is reported that numerous models or real networks incorporate the linear preferential attachment [12－13].

But where does the preferential attachment come from? Recently, several papers have offered promising proposals and models that shed some light on this issue [14, 15], but a universal answer is still lacking. In this paper, we try to give answer to this question through studying the structure of complex networks from another point of view － statistical mechanics, which is a well-founded general theory with true predictive power. We put the case of sparse limit as keystone in our work, and reveal the structure of growing networks with linear preferential attachment by the special connection probability. The BA model [7], as a highly influential model in the complex network field and typically has a linear preferential attachment, deserves our notice for its topological structure when we inspect it from the novel view.

In particular, we develop a method using the connection function to calculate the degree correlation and clustering coefficient in our ensemble model. These analytical predictions are tested by numerical simulation in BA model. Other models or real networks incorporating the sparse and linear preferential attachment characteristics can also be studied by using this method.

# 2. Partition function and connection probability

We investigate the structure of complex networks, basing on the exponential random model [16,17], which are a class of graph ensembles of fixed vertex number defined by analogy with the Boltzmann ensemble of statistical mechanics and can be derived from the first principle of maximum entropy. Considering that there is one link at most between each vertex pair, we map the topological structure of complex network into the equilibrium Fermi gas, with each vertex pair corresponding to a special state of Fermi gas and the number of edges between vertex pair representing the number of particle on the state. As the degree sequence $\{k_i\}$ is a set of microcosmic quantities, it can be used to explore the microcosmic structure of a network. According to the exponential random graph model [18, 19], network with this set of observations have Hamiltonian, $H = \sum_i q_i k_i$, where the parameter $q_i$ for vertex $i$ is similar to the inverse-temperature or field parameter in equilibrium statistical mechanics [20,21]. Taking into account that $k_i = \sum_j s_{ij}$, where $s_{ij}$ is the element of the adjacent matrix whose value is 1 if there is a link between vertex pair $(i, j)$ and 0 otherwise, the Hamiltonian can be written as $H = \sum_{i<j}(q_i + q_j)s_{ij} = \sum_{i<j} q_{ij} s_{ij}$, where $q_{ij}$ denotes $(q_i + q_j)$, and then the partition function of the system is

$$Z = \sum_{\{s_{ij}\}} \exp(-\sum_{i<j} q_{ij} s_{ij}) = \prod_{i<j} \sum_{s_{ij}=0}^{1} \exp(-q_{ij} s_{ij}) = \prod_{i<j}(1 + e^{-q_{ij}}). \tag{1}$$

The free energy is $F = -\ln Z$, and the connection probability $p_{ij}$ between $i$ and $j$ can be given by

$$p_{ij} = <s_{ij}> = \frac{\partial F}{\partial q_{ij}} = -\frac{\partial \ln Z}{\partial q_{ij}} = \frac{1}{e^{q_{ij}}+1} = \frac{1}{e^{q_i+q_j}+1}. \tag{2}$$

Similarly, the expectation value of the degree of vertex $i$ is

$$\bar{k}_i = \frac{\partial F}{\partial q_i} = -\frac{\partial \ln Z}{\partial q_i} = \sum_j \frac{1}{e^{q_i+q_j}+1} = \sum_j p_{ij}. \tag{3}$$

In the sparse limit, $p_{ij} \ll 1$, Eq.(2) can be written as

$$p_{ij} = e^{-q_i - q_j}. \tag{4}$$

In this case, the network system corresponds to the non-interacting ideal gas, and the connection probability corresponds to the Maxwell-Boltzmann distribution in statistical mechanics [22]. Thus Eq. (3) turns into

$$\bar{k}_i = \sum_j p_{ij} = e^{-q_i} \sum_j e^{-q_j} = A e^{-q_i}, \tag{5}$$

where $A \equiv \sum_j e^{-q_j}$. Therefore, the expected degree $\bar{k}_i$ within our graph ensemble is proportional to the factor $e^{-q_i}$. Noting that the expected (or average) total number of links is $\bar{l} = \frac{1}{2}\sum_i \bar{k}_i$, we can get $A = \sqrt{2\bar{l}}$, hence

$$p_{ij} = e^{-q_i} e^{-q_j} = \frac{\bar{k}_i}{\sqrt{2\bar{l}}} \cdot \frac{\bar{k}_j}{\sqrt{2\bar{l}}} = \frac{\bar{k}_i \bar{k}_j}{2\bar{l}}. \tag{6}$$

Thus we derived the connection probability from the basic assumption of our model.

## 3. Comparison with the CL model and BA model

The connection probability (6) is proportional to the product of the degrees of its endpoints, which is precisely the model of Chung and Lu (CL model) [23], but is obtained under the approximation (4). We refer to such case as the "sparse limit" or the "classical limit" corresponding to the phenomenon in quantum gases at low density. Therefore, the model of Chung and Lu is available and appropriate for the sparse networks as it maximizes the entropy of the system and occurs with the largest probability. In addition, the conditional probability $\prod_i$ that vertex $i$ will be connected to vertex $j$ is

$$\prod_i(k_j) = \frac{p_{ij}}{\sum_m p_{im}} = \frac{\overline{k}_j}{\sum_m \overline{k}_m}, \quad (7)$$

which is only dependent on the expected degree $\overline{k}_j$ of vertex $j$, indicating that the connection probability of our model embodies the linear preferential attachment. Conversely, if we adopt the linear preferential attachment (7) as done in BA model, we can get the connection probability (6).

In the complex network field, various preferential attachments are used in the growing models and different preferential attachment leads to different exponents of the scale free behavior [8－10]. For instance, in the BA model [7], the linear preferential attachment leads to a power law degree distribution with exponent –3, while in other models the preferential attachment is adjusted to obtain other exponents that agree more closely to real networks. However, the connection function (6) is based on the maximum entropy principle, implying that such structure happens with the largest probability.

Let us apply the above analysis to the BA model, in which a new vertex is added and linked to $m$ present vertices with linear preferential attachment according to their degrees. At time $t$, select two vertices $i$ and $j$, which were added at time $t_i$ and $t_j$, their average degrees are given by $\overline{k}_i(t) = m(\frac{t}{t_i})^{0.5}$ and $\overline{k}_j(t) = m(\frac{t}{t_j})^{0.5}$ [7]. According to the connection probability (6), we have

$$p_{ij} = \frac{\overline{k}_i(t)\overline{k}_j(t)}{\sum_n \overline{k}_n(t)} = \frac{m^2 t(t_i t_j)^{-0.5}}{2mt} = \frac{m}{2}(\frac{1}{t_i t_j})^{0.5}. \quad (8)$$

Strikingly, the connection probability is independent of time, indicating that the vertices added later do not influence links established earlier. As a result, the BA model is also a model of Chung and Lu despite its dynamical and evolutional behaviors. The growing mechanism of BA model keeps the sparseness of networks, and the linear preferential attachment based on growth keeps the maximal entropy of the system. This highlights an alternative microscopic mechanism accounting for the presence of the linear preferential attachment dynamics in growing networks. In this sense, it is reasonable to speculate that growing network models based on growth and linear preferential attachment are the normal state of sparse networks.

## 4. Degree correlation and clustering coefficient

Let us use the connection probability (6) to calculate the ensemble mean of the degree correlation and clustering coefficient. The degree correlation can be quantified by calculating the mean degree of the network neighbors of a vertex as a function of the degree $k$ of that vertex [24]. The mean sum of the degrees of the neighbors of a vertex $i$, which is denoted by $\bar{K}_i^{nn}$, can be given by $\bar{K}_i^{nn} = \sum_j p_{ij}(\bar{k}_j + 1) = \dfrac{\bar{k}_i \sum_j \bar{k}_j(\bar{k}_j + 1)}{\sum_j \bar{k}_j}$ [25]. Thus the mean degree of a neighbor of vertex $i$ is equal to

$$\bar{k}_i^{nn} = \frac{\bar{K}_i^{nn}}{\bar{k}_i} = \frac{\sum_m \bar{k}_m^2}{\sum_m \bar{k}_m} + 1 = \frac{<\bar{k}^2>}{<\bar{k}>} + 1. \tag{9}$$

Throughout this paper, we denote the expected values or ensemble mean by a bar (e.g., $\bar{k}$), and the average over network vertices by a bracket (e.g., $<\bar{k}>$) while the actual values in a particular graph by undecorated characters. For the BA model, at time $t$, the minimal and maximal degree are given by $\bar{k}_{\min}(t) = m$ and $\bar{k}_{\max}(t) = mt^{0.5}$. Combining $P(k) = \dfrac{2m^2}{k^3}$, we have

$$\bar{k}_i^{nn}(t) = \frac{m \ln t}{2} + 1. \tag{10}$$

Eq. (10) is independent of $\bar{k}_i(t)$, indicating that the degrees of connected vertices are uncorrelated. Therefore, the degree correlation coefficient defined by Newman [26] is equal to 0 for the BA model. The form of formula (10) indicates that the mean degree of neighbors $\bar{k}_i^{nn}(t)$ increases logarithmically with the network size $t$.

Next, we derive analytical formula for the clustering coefficient. Before we get the clustering coefficient of a certain vertex $i$, we need to calculate the ensemble mean of edges among its nearest neighbors. Consider two arbitrary vertices $m$ and $n$. The probability that these two vertices connecting to vertex $i$ simultaneously can be given by $p_{im} p_{in}^{im}$, where $p_{in}^{im}$ is the connection probability between $i$ and $n$ on the condition that $i$ and $m$ are already connected. In the same way, we use $p_{mn}^{im,in}$ to denote the connection probability between $m$ and $n$, provided that they have a common vertex $i$. The average connections $\bar{E}_i$ between the nearest neighbors of vertex $i$, can be formally computed as the probability that vertex $i$ is connected to vertices $m$ and $n$, and that those two vertices are at the same time joined by an edge, summing over all the possible vertex pairs $m$ and $n$, and divided by 2. Thus, we can write the clustering coefficient of vertex $i$ as

$$\bar{C}_i = \frac{\bar{E}_i}{\bar{k}_i(\bar{k}_i-1)/2} = \frac{2\bar{E}_i}{\bar{k}_i(\bar{k}_i-1)} = \frac{\sum_{m,n} p_{im} p_{in}^{im} p_{mn}^{im,in}}{\bar{k}_i(\bar{k}_i-1)} \qquad (11)$$

In our ensemble model, each link is present or absent independently of all others, therefore the conditional probability can be simplified as $p_{in}^{im} = p_{in} = \frac{\bar{k}_i \bar{k}_n}{2\bar{l}}$, $p_{mn}^{im,in} = p_{mn} = \frac{\bar{k}_m \bar{k}_n}{2\bar{l}}$. Substituting these formulas into (11) and assuming $\bar{k}_i$ is large enough, we get

$$\bar{C}_i \approx \frac{\sum_{m,n} p_{im} p_{in} p_{mn}}{\bar{k}_i^2} = \frac{(\sum_i \bar{k}_i^2)^2}{(2\bar{l})^3} = \frac{(<\bar{k}^2>)^2}{N<\bar{k}>^3}. \qquad (12)$$

where $N$ is the network size or number of vertex. This result is close to the result in Ref. [27] for large expected degrees. Noting that $P(k) = 2m^2/k^3$ in BA model, we have

$$\bar{C}_i = \frac{m \ln^2 t}{8t}. \qquad (13)$$

Eq. (13) is consistent with the results in Ref. [28], in which the rate equation approach is used to calculate the two-vertex and three-vertex correlations for the degree correlation and clustering coefficient. This formula shows that the clustering coefficient of each vertex in the BA model is uniform and only depends on the network size induced by the addition of new vertices. As $t$ grows, the clustering coefficient scales as $\ln^2 t/t$.

In order to check the analytical predictions, we performed numerical simulations for the BA model with different values of $m$, and varying network size $t$. The main results are shown in Fig.1. As expected, the numerical predictions of the mean degree of neighbors $\bar{k}^{nn}$ and clustering coefficient $\bar{C}$ show very good agreement with the analytical results $\bar{k}^{nn}(t) \sim \ln t$ and $\bar{C}(t) \sim \ln^2 t/t$.

We would like to stress that the above analysis method is suitable for various sparse networks including the BA model and real sparse networks. Considering that the growing network models with linear preferential attachment has a connection probability of Eq. (6), we can give analogous analyses of the degree correlation and clustering coefficient to other growing network models, such as models proposed in Ref. [29, 30].

## 5. Discussions and conclusions

In this paper we investigate networks from the first principle of maximum entropy borrowed from the statistical mechanics, and derive analytical expressions for the connection probability in the sparse limit. Thus we provide a novel and convincing mechanism for the ubiquitous linear preferential attachment in scale-free networks. As has been mentioned in the introduction, social systems, biologic systems, and other systems that can be described by network display the linear preferential attachment behavior, and therefore are all in a maximal entropy state.

Through the analysis carried out in this paper we uncover the relation between BA model and CL model. We emphasize the importance of connection probability, and demonstrate that many characteristics including the degree correlation and clustering coefficient can be calculated by the connection function in our ensemble model. Especially, in the sparse limit, the expected degrees are uncorrelated and the clustering coefficient is uniform，in contrary to the non-sparse case, where the degrees are correlated and clustering coefficient is heterogeneous[31,27].

It is worth mentioning that all the quantities in this paper, has been derived on the basis of a canonical ensemble in which the fixed quantity is the expected degree sequence or average degree sequence [32]. Properties of random graphs in such an ensemble strongly depend on the expected degree sequence. However, it is impossible to find out the expected degree sequence from empirical data. In practice, it is often the case that we only measure the degree sequence once, for instance, we have only one Internet, and hence only one measurement of the degree sequence. The observed degree sequence $\{k_i\}$ fluctuates around the expected degree sequence $\{\bar{k}_i\}$. In the sparse limit, the degree fluctuation of vertex $i$ is given by $\overline{k_i^2} - (\bar{k}_i)^2 = -\frac{\partial^2 F}{\partial q_i^2} = \bar{k}_i$. The magnitude of fluctuation around the average is reduced for large $\bar{k}_i$ since $\sqrt{\overline{k_i^2} - (\bar{k}_i)^2}/\bar{k}_i = 1/\sqrt{\bar{k}_i}$. Therefore, we expect the observed degree $k_i$ is close to the average degree $\bar{k}_i$ for large $\bar{k}_i$. In this case, our best estimate of the expected degree sequence is approximately equal to the one measurement that we have. Thus, we can use the observed degree sequence in real sparse networks to estimate the ensemble mean of degree correlation and clustering coefficient. For this purpose, we suggest a method to construct an ensemble, in which the expected degrees are equal to the observed degrees.

Our construction method is as follows. Once the degree sequence in a real sparse network is measured, we place $l = \frac{1}{2}\sum_i k_i$ edges between vertex pairs $(i,j)$ with probability $p_{ij} = \frac{k_i k_j}{2l}$ as a realization of network in the ensemble. In order to minimize the statistical fluctuations, we average the physical quantities over a large number of realizations. This construction method is suitable for sparse networks with various degree sequences.

The analytic techniques of equilibrium statistical mechanics are ideally suitable for the study of network models and can shed much light on their structure and behavior. The approach adopted in this paper will be useful in the future research.

**Acknowledgement** This work is supported by the Ministry of Education of China under project 704035 and by NSFC under projects 10375025, 10275027.

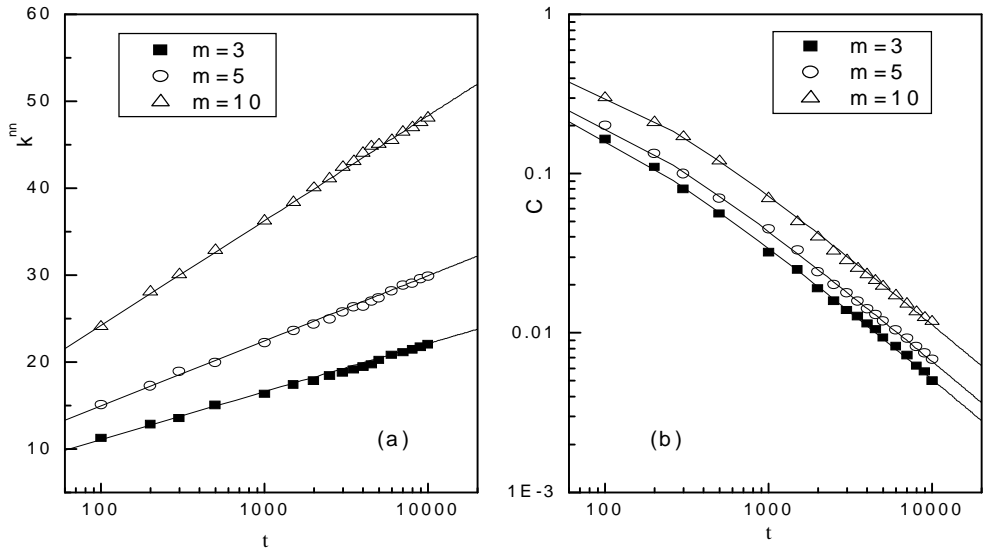

FIG.1. (a) Average nearest neighbor degree $\bar{k}_i^{nn}(t)$ as a function of the network size $t$ for different values of $m$. (b) Clustering coefficient versus network size $t$ for various values of $m$. The solid lines are least-squares fits to the form $\bar{k}^{nn}(t) \sim \ln t$ in (a) and $\bar{C}(t) \sim \ln^2 t / t$ in (b), as predicted by Eq. (10) and (13).